# Competing effects of surface phonon softening and quantum size effects on the superconducting properties of nanostructured Pb

Sangita Bose,[1*#] Charudatta Galande, S. P. Chockalingam, Rajarshi Banerjee,[2] Pratap Raychaudhuri,[†1] and Pushan Ayyub[‡1]

[1]Department of Condensed Matter Physics and Material Science, Tata Institute of Fundamental Research, Mumbai 400005, India

[2]Department of Materials Science and Engineering, University of North Texas, Denton, Texas 76203-5310, U.S.A.

## Abstract

The superconducting transition temperature ($T_C$) in nanostructured Pb remains nearly constant as the particle size is reduced from 65 to 7nm, below which size the superconductivity is lost rather abruptly. In contrast, there is a large enhancement in the upper critical field ($H_{C2}$) in the same size regime. We explore the origin of the unusual robustness of the $T_C$ over such a large particle size range in nanostructured Pb, by measuring the temperature dependence of the superconducting energy gap in planar tunnel junctions of $Al/Al_2O_3$/nano-Pb. We show that below 22nm, the electron phonon coupling strength increases monotonically with decreasing particle size, and almost exactly compensates for the quantum size effect, which is expected to suppress $T_C$.

---

[#] Present address: Max Planck Institute for Solid state Research, Nanoscale Science Department, Stuttgart, Germany.

[*] Electronic Mail: Sangita.Bose@fkf.mpg.de

[†] Electronic Mail: pratap@tifr.res.in

[‡] Electronic Mail: pushan@tifr.res.in

Superconductivity at reduced length scales has been a subject of intense research over the past few decades.[1,2,3,4,5,6,7,8,9,10,11] Though one may expect changes in the superconducting properties as the system size is reduced below the fundamental length scales such as the coherence length, $\xi(T)$, and the penetration depth, $\lambda_L(T)$, it is now established that there is actually a *third* length scale that finally defines a zero dimensional superconductor. This is the critical particle diameter ($D_C$) at which the energy level spacing ($\delta$) arising from the discretization of the energy bands (the "Kubo" gap) equals the superconducting energy gap ($\Delta(0)$). Superconductivity is completely destabilized below this length scale. The existence of such an 'Anderson criterion'[2] has been successfully demonstrated in many elemental superconductors such as Al,[3] Sn,[5] In,[12] Pb,[6,7] and Nb.[13] However, as the size of the superconductor approaches $D_C$, the behavior of the superconducting transition temperature ($T_C$) is quite different in different systems: superconductors with a weak electron phonon coupling (In, Al, and Sn) show an increase in $T_C$; the intermediate coupling superconductor Nb shows a gradual, monotonic decrease in $T_C$; while the $T_C$ in the strong coupling superconductor, Pb, shows almost no change.

Two competing mechanisms control the $T_C$ in nanostructured superconductors. The first arises from the increase in surface to volume ratio with decreasing size. As the surface atoms have a smaller coordination number than the bulk atoms, surface phonons are softer than bulk phonons. This leads to an overall decrease in the phonon frequencies in nanoparticles,[14] resulting in an enhanced electron-phonon coupling strength[15] and a higher $T_C$. Experimentally, an increase in the electron-phonon coupling can be detected by measuring the dimensionless quantity: $2\Delta(0)/k_BT_C$, which monotonically increases with coupling strength from its value of 3.52 in the weak coupling limit. This effect could be counteracted by the quantum size effect (QSE) arising from the discretization of the electronic energy bands in small particles and leading to a decrease in the effective density of states, $N(0)$, at the Fermi level.[16,17] However, within the Bardeen-Cooper-Shrieffer theory, both $\Delta(0)$ and $T_C$ are related to $N(0)V$ by the same functional form ($V$ is the effective electron phonon interaction potential). QSEs, therefore, do not lead to a change in $2\Delta(0)/k_BT_C$ with reduction in particle size. To distinguish between the effects of these two mechanisms in nanostructured superconductors, it is necessary to make independent measurements of $T_C$ and $\Delta(0)$ as a function of particle size.

We have earlier shown[13] that in nanocrystalline thin films of Nb, QSEs become apparent below ≈ 20nm. The QSE-induced reduction in the density of states at the Fermi level decreases $T_C$ to almost 50% of its bulk value as the particle size is reduced from 20 to 8nm. In the strong coupling superconductor Pb, the bulk superconducting energy gap ($\Delta(0) \approx 1.38$meV) as well as the critical size at which superconductivity gets destroyed ($D_C \approx 6$nm) are close to the corresponding values in Nb. One may therefore expect QSE to play similar roles in Pb and Nb. The size dependence of $T_C$ in Pb is, however, qualitatively different from that in Nb, decreasing by only by ≈13% as particle size is reduced[6,7] from bulk to 7nm, below which it becomes non-superconducting. To understand the robustness of the $T_C$ with decreasing size in Pb, we carried out simultaneous measurements of $\Delta(0)$ and $T_C$ in planar tunnel junctions consisting of Al, $Al_2O_3$, and nanostructured Pb films (with different values of $D$), grown by high pressure magnetron sputter deposition. Interestingly, $\Delta(0)$ was found to increase with decreasing size (for $D$ < 20nm) though $T_C$ remained virtually constant. A measurement of the temperature variation of the gap further indicates a size dependent deviation from the weak coupling BCS behavior in nanostructured Pb, implying an enhancement of the electron phonon coupling strength. Our results suggest that in nano-Pb, the decrease in $T_C$ due to quantum size effect is almost exactly offset by the increase in electron phonon coupling strength, down to the Anderson limit.

Nanocrystalline films of Pb (≈200nm thick) were deposited on glass substrates by high-pressure magnetron sputtering from elemental Pb targets (Kurt and Lesker, 99.999%). The particle size was varied in the range 5-60nm by controlling the sputtering gas (Ar) pressure, the applied power and the deposition time. To prevent oxidation, the nanocrystalline Pb films were capped with a 40nm thick overlayer of Si grown in-situ using RF-supttering. The mean particle size ($D$) and size distribution were determined from x-ray diffraction (XRD) line profile analysis using WINFIT software and transmission electron microscopy (TEM). The particle sizes measured by the two methods matched closely. The particle size distribution in each film was approximately ±15%. Figure 1(a) and 1(b) show the bright field TEM patterns obtained from samples with $D_{XRD}$ = 60 and 14nm, respectively, the insets showing the corresponding selected area diffraction patterns. The high resolution TEM image of the $D_{XRD}$ = 5nm sample (Fig. 1(c)) shows crystalline Pb grains of ≈5nm diameter (darker contrast) separated by a disordered intergranular region, possibly consisting of an amorphous Pb-O phase.[18] The Pb grains are electronically isolated to some degree by the intergranular region. Magnetization and transport measurements were carried out down to 2.2 K using a planar coil ac susceptometer and a magneto-transport set up, respectively. The superconducting energy gap (Δ) was measured by fabricating planar tunnel junctions of $Al/Al_2O_3$/nano-Pb films by a standard method.[19] 1mm Al strips were thermally evaporated through a mask on a glass side (thoroughly cleaned in boiling acetone and vapor cleaned in trichloroethylene and propanol). The Al film was oxidized by exposing to air for 20 minutes. Nanostructured Pb films were then sputter deposited as cross strips on the $Al_2O_3$ layer using proper masks to produce on each device two $Al/Al_2O_3$/nano-Pb tunnel junctions with effective junction areas of 1×1 $mm^2$. For the tunnel junction with largest particle size ($D_{XRD}$=64nm) the Pb strip was thermally evaporated at high vacuum.

Figures 2(a) and (b) show the temperature dependence of the ac magnetic susceptibility and the dc resistivity for the nano-Pb films with different particle size. Using commonly accepted criteria, $T_C$ was estimated (i) from transport measurements as the temperature at which the resistance dropped to 10% of its normal state value, and (ii) from ac susceptibility measurements as the temperature at which the real part of susceptibility deviated from zero. The $T_C$ obtained from the two methods matched almost exactly (Fig. 3(a)). We observe that the $T_C$ does not deviate from the bulk value (7.24K) down to $D \approx$ 14nm, and decreases by only ≈13% between 14 and 7nm (Fig. 3(a)). Below 7nm, Pb loses superconductivity. The calculated critical particle diameter (Anderson limit) is 6nm.[2] This result agrees with the critical size reported by other groups.[6, 7] Fig. 3(a) also brings out the qualitative difference in the size dependence of $T_C$ in nano-Pb and nano-Nb (data from Ref. 13). In the intermediate coupling superconductor, Nb, a gradual depression of $T_C$ starts at comparatively large sizes (above 20nm), and the $T_C$ decreases by about 50% down to 8nm, below which it becomes non-superconducting. Figure 3(b) shows the upper critical field measured at 4.2K. The $H_{C2}$ measured at 4.2K, shows a monotonic increase with decreasing size down to 7nm, consistent with previous results[7]. In nano-Nb, $H_{C2}$ shows a non-monotonic size dependence[20], with an increase down to 20nm and steadily decreasing at lower sizes.

The evolution of the superconducting gap (Δ) with both size and temperature was obtained from 4-probe *I-V* measurements down to 2.2K, in planar tunnel junctions of Al/$Al_2O_3$/nano-Pb. Figures 4(a) and (b) show plots of the differential conductance, $G(V)$ = d$I$/d$V$, versus voltage for the samples with $D_{XRD} \approx$ 64nm and 11nm, respectively. The data was normalized with respect to the conductance values ($G_n$) between 3mV and 4mV to avoid the phonon contribution that occur at higher bias. The tunneling spectra (*G* vs. *V*) were fitted to a theoretical model for tunneling between a superconductor and a normal metal. The ±15% distribution in particle size leads to a corresponding distribution in Δ, which is accounted for by broadening the BCS density of states[21] with a linewidth Γ. The broadened density of states is given by:[22] $N(E,\Gamma) = \mathrm{Re}\left(\frac{E+i\Gamma}{\sqrt{(E+i\Gamma)^2-\Delta^2}}\right)$. Thus, there are two fitting parameters, Δ and Γ, for each temperature. The gap, obtained by fitting the tunneling spectra, increases from its bulk value[23], Δ(0) ≈ 1.38 meV for the film with $D_{XRD}$ = 64nm to Δ(0) ≈ 1.88 meV for $D_{XRD}$ = 9nm ($T_C$ = 6.9K). Correspondingly, the coupling strength $2\Delta(0)/k_BT_C$ increases from 4.37 for the 64nm sample to 6.28 at 9nm (Fig. 4(c)). In contrast, in nanostructured Nb, $2\Delta(0)/k_BT_C$ remains close to the BCS value[13] in the same size range. The size-dependent increase in the coupling strength can be independently inferred from the temperature variation of the superconducting gap (Fig. 4(d)). Since bulk Pb is a strong coupling superconductor, the normalized value of Δ deviates slightly from the weak coupling BCS curve. However, this deviation progressively increases towards the strong coupling limit as the particle size is reduced. Thus, Pb becomes a stronger coupling superconductor in the nanocrystalline state.

We can also explain the monotonic increase in $H_{C2}$ down to 7nm within the above scenario. The upper critical field is given by: $H_{C2} = \frac{\phi_0}{2\pi(\xi_0 l_{eff})^2}$, where $\phi_0$ is the flux quantum, $\xi_0$ [$\propto 1/T_c N(0)$] is the Pippard coherence length and $l_{eff}$, the electronic mean free path. The observed increase in $H_{C2}$ with size reduction indicates that the decrease in the effective mean free path due to higher fraction of grain boundaries in the nanostructured system, overrides the effect of the decrease in the product [$T_C N(0)$] that would arise from QSE.

The increase in the electron-phonon coupling strength with decreasing size, inferred from the temperature dependence of the superconducting gap, implies that phonon softening plays an important role in influencing $T_C$ in this strong coupling superconductor. The weak coupling BCS equations for $T_C$ are thus no longer valid in the nanostructured system. In the strong coupling limit, $T_C$ is given by McMillan's equation:[15] $T_C = \frac{\Theta_D}{1.45} \exp\left[\frac{-1.04(1+\lambda)}{\lambda - \mu^*(1+0.62\lambda)}\right]$, where $\Theta_D$ is the Debye temperature and $\mu^*$ is the effective electron-electron repulsion term. McMillan also showed that the $\lambda$ is approximately proportional to the inverse of the average squared phonon frequency: $\langle\omega^2\rangle_M = \langle\omega\rangle/\langle 1/\omega\rangle$. Thus, a reduction in $\langle\omega^2\rangle_M$, should lead to an increase in $\lambda$. It is clear from McMillan's equation that a higher coupling strength (due to size reduction) should produce higher $T_C$. In fact, the enhanced $T_C$ observed in nanostructured, weak coupling superconductors such as Al, Sn, Ga, and In are ascribed to phonon softening due to surface effects. Why do we not observe a similar increase in $T_C$ in the strong coupling superconductor Pb? A plausible explanation for the observed size independence of $T_C$ is the cancellation of the opposing influences of phonon softening and QSE. The quantization of the electronic wave vector arising from the discretization of the energy levels at small sizes cannot be neglected since it leads to the Anderson criterion that correctly predicts the observed destabilization of superconductivity at 6nm in Pb. Also, from our measurements of $T_C$ and $\Delta$ in nanostructured Nb, we know that the QSE plays a dominant role[13] in influencing $T_C$ by decreasing the density of states at the Fermi level in small particles. We therefore believe that phonon softening effects (that tend to increase $T_C$) are almost exactly offset by QSE (that tend to decrease $T_C$) in Pb down to 14nm, resulting in a size invariant $T_C$. Between 14 and 7nm, QSE dominates and produces a 13% decrease in $T_C$. Below 7nm, Pb is no longer superconducting.

In summary, we have reported a detailed investigation of the evolution of superconductivity with particle size in nanostructured films of Pb. Measurements of the superconducting energy gap and the critical fields in nanostructured Pb indicate a deviation from the weak coupling BCS behavior. This is indicative of an increase in the electron-phonon coupling strength with reduction in particle size in nano-Pb. The expected increase in $T_C$ due to this effect is however not observed, being almost exactly offset by the quantum size effect. Below 7nm, quantum size effects dominate and superconductivity is destroyed, consistent with the Anderson criterion. Our studies provide a natural explanation for the robustness of the superconducting transition temperature in nanosized Pb particles down to the theoretical limit at which superconductivity is destabilized.

*Acknowledgement:* We thank J. John, N. Kulkarni and V. Bagwe for technical assistance.

## Figure Captions

**Figure 1.** Bright field TEM images of the nanostructured Pb films with particle size (a) $D_{XRD}$ = 60nm and (b) $D_{XRD}$ = 14nm, the corresponding selected area diffraction patterns being shown in the insets. (c) High resolution TEM image of the film with $D_{XRD}$ = 5nm showing Pb grains (marked by circles) separated by a disordered intergranular region.

**Figure 2.** Temperature dependence of the (a) normalized ac susceptibility (b) dc electrical resistance of the nanostructured Pb films with different particle sizes.

**Figure 3.** (a) Particle size ($D$) dependence of the superconducting transition temperature ($T_C$) of Pb obtained from electrical transport (circles) and magnetic susceptibility (triangles). The size dependence of $T_C$ for Nb (diamonds) (from ref. 13) is shown for comparison (squares). (b) Variation of the upper critical field ($H_{C2}$) with particle size obtained from magnetoresistance data.

**Figure 4.** Tunnelling spectra (normalized differential conductance vs. applied voltage) recorded at different temperatures for the nanostructured Pb films with (a) $D_{XRD}$ = 64 nm, and (b) $D_{XRD}$ = 11nm. The symbols denote the experimental points and the solid lines are the theoretical fits. (c) Size dependence of the coupling strength, $2\Delta(0)/k_BT_C$, for Pb and Nb (Ref. 13). (d) Variation of the normalized superconducting energy gap with reduced temperature for Pb films with $D_{XRD}$ = 64nm, 22nm and 11nm. The solid line is the temperature variation of the gap obtained from the weak coupling BCS equation; *(inset)* an expanded portion of the same data showing more clearly the increasing deviation from weak coupling behaviour close to $T_c$.

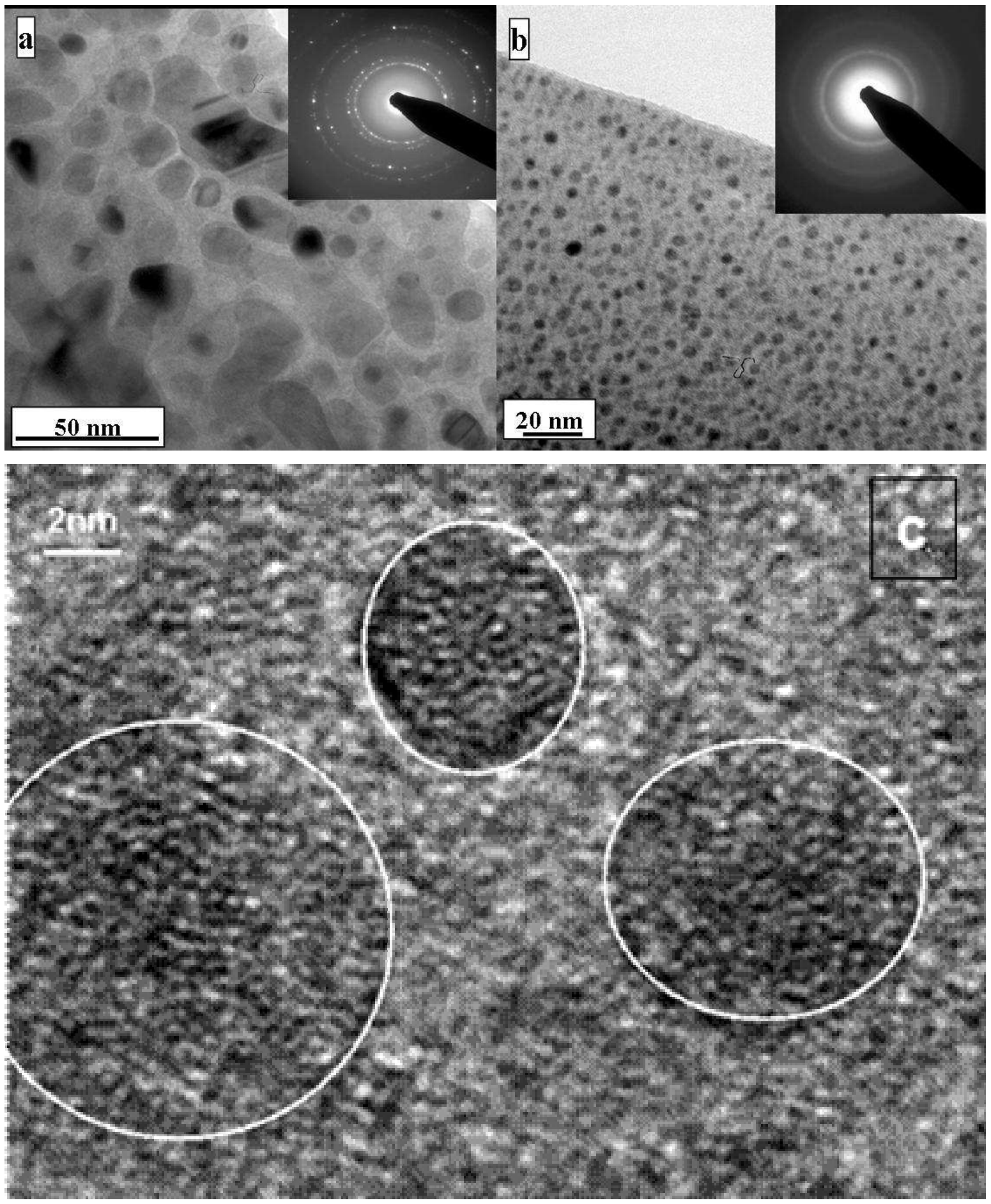

Figure 1

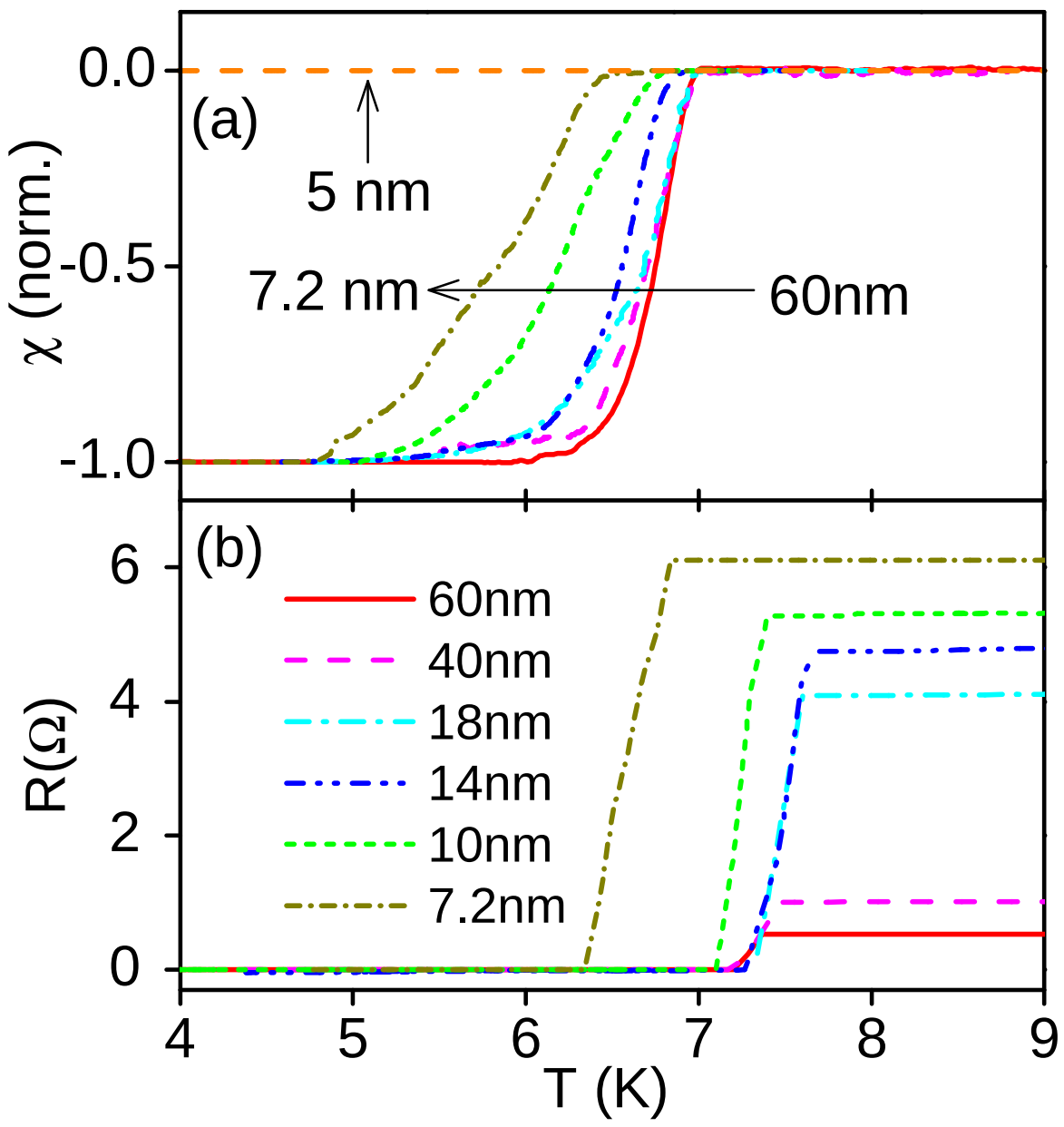

Figure 2

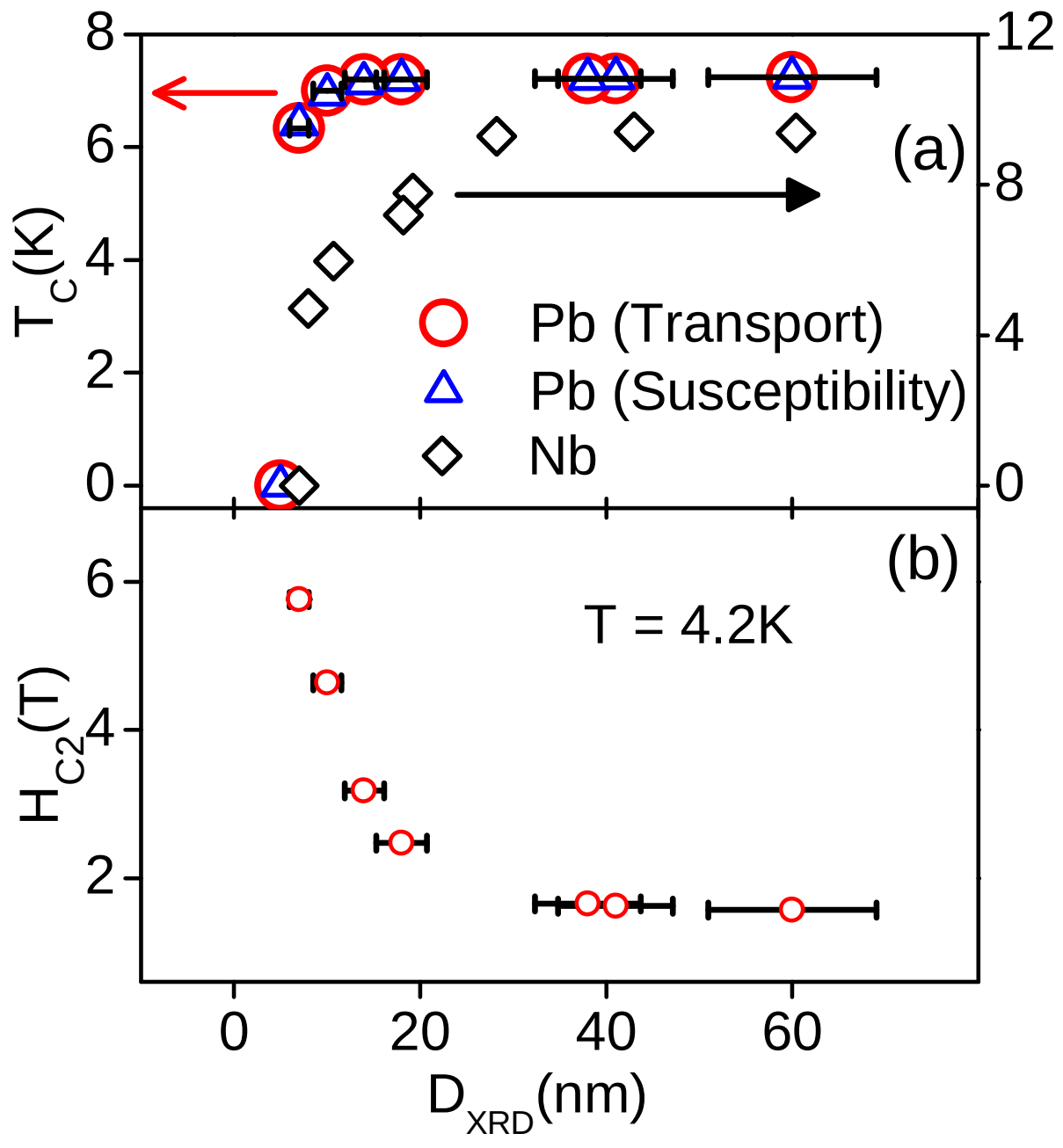

Figure 3

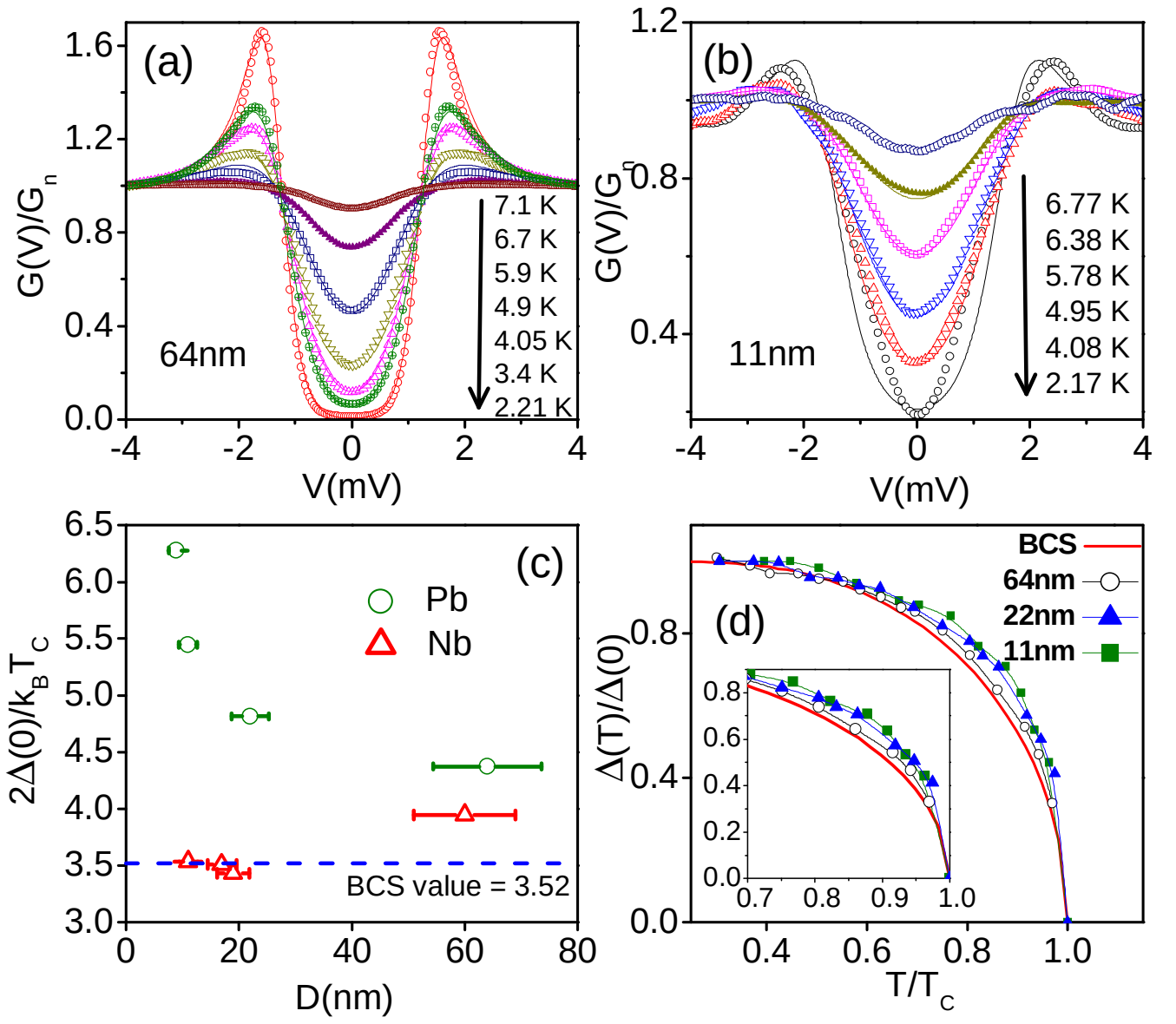

Figure 4

## References

[1] S. Matsuo, H. Sugiura, and S. Noguchi, J. Low Temp. Phys. **15**, 481 (1974).

[2] P.W. Anderson, J. Phys. Chem. Solids **11**, 26 (1959).

[3] K. Oshima, T. Kuroishi, T. Fujita, J. Phys. Soc. Jpn. **41**, 1234 (1976).

[4] B. Abeles, R. W. Cohen, G. W. Cullen, Phys. Rev. Lett. **17**, 632 (1966).

[5] N.A.H.K. Rao, J. C. Garland, D. B. Tanner, Phys. Rev. B **29**, 1214 (1984); T. Tsuboi and T. Suzuki, J. Phys. Soc. Jpn. **42**, 437 (1977).

[6] S. Reich, G. Leitus, R. Popovitz-Biro, M. Schechter, Phys. Rev. Lett. **91,** 147001-1(2003)**.**

[7] W. H. Li, C.C.Yang, F. C. Tsao, K. C. Lee, Phys. Rev. B **68**, 184507 (2003).

[8] H. M. Jaeger, D. B. Haviland, B. G. Orr, A. M. Goldman, Phys. Rev. B **40**, 182 (1989), and references therein.

[9] T. Hihara, Y. Yamada, M. Katoh, D. L. Peng, K. Sumiyama, J. Appl. Phys. **94**, 7594 (2003).

[10] R. P. Barber, Jr, L. M. Merchant, A. La Porta, R. C. Dynes, Phys. Rev. B **49**, 3409 (1994); A. Frydman, O. Naaman, and R. C. Dynes, Phys. Rev. B **66**, 052509 (2002)

[11] D.C. Ralph, C.T. Black, and M. Tinkham, Phys. Rev. Lett. **74**, 3241 (1995).

[12] W.-H. Li, C. C. Yang, F. C. Tsao, S. Y. Wu, P. J. Huang, M. K. Chung, and Y. D. Yao, Phys. Rev. B, 72, 214516 (2005)

[13] S. Bose, P. Raychaudhuri, R. Banerjee, P. Vasa, and P. Ayyub, Phys. Rev. Lett. **95**, 147003 (2005)

[14] J. M. Dickey, A. Paskin, Phys. Rev. Lett. **21**, 1441 (1968).

[15] W. L. McMillan, Phys. Rev. **167**, 331 (1968).

[16] M. Strongin, R. S. Thompson, O. F. Kammerer, and J. E. Crow, Phys. Rev. B **1**, 1078 (1970).

[17] M. Strongin, O. F. Kammerer, J. E. Crow, R. D. Parks, D. H. Douglass, M. A. Jensen, Phys. Rev. Lett. **21**, 1320 (1968)

[18] Similar structure was also observed in nanocrystalline Nb films where electron energy loss spectroscopy confirmed the granular and intergranular region to be Nb and $Nb_2O_5$ respectively, see S. Bose, R. Banerjee, A, Genc, P. Raychaudhuri, H. L Fraser and P. Ayyub, J. Phys.: Condens. Matter **18**, 4553 (2006).

[19] I. Giaever, Phys. Rev. Lett. **5**, 147 (1960); I. Giaever, Phys. Rev. Lett. **5**, 464 (1960)

[20] Sangita Bose, Pratap Raychaudhuri, Rajarshi Banerjee, and Pushan Ayyub, Phys. Rev. B **74**, 224502 (2006).

[21] It has been shown in the context of anisotropic superconductors that the conductance spectrum arising from a distribution of superconducting energy gaps can be fitted by incorporating the broadening parameter Γ: P Raychaudhuri, D Jaiswal-Nagar, Goutam Sheet, S Ramakrishnan and H Takeya, Phys. Rev. Lett. **93,** 156802 (2004).

[22] R. C. Dynes, J. P. Garno, G. B. Hertel and T. P. Orlando, Phys. Rev. Lett. **53**, 2437 (1984)

[23] Ivar Giaever and Karl Megerle, Phys. Rev. **122**, 1101 (1961); P. Townsend and J. Sutton, Phys. Rev. **128**, 591 (1962).